# Modeling Cyber-Physical Systems: Model-Driven Specification of Energy Efficient Buildings

Thomas Kurpick
Software Engineering
RWTH Aachen University
Aachen, Germany
http://se-rwth.de

Markus Look
Software Engineering
RWTH Aachen University
Aachen, Germany
http://se-rwth.de

Claas Pinkernell
synavision GmbH
Aachen, Germany
http://synavision.de

Bernhard Rumpe
Software Engineering
RWTH Aachen University
Aachen, Germany
http://se-rwth.de

## ABSTRACT
A lot of current buildings are operated energy inefficient and offer a great potential to reduce the overall energy consumption and $CO_2$ emission. Detecting these inefficiencies is a complicated task and needs domain experts that are able to identify them. Most approaches try to support detection by focussing on monitoring the building's operation and visualizing data. Instead our approach focuses on using techniques taken from the cyber-physical systems' modeling domain. We create a model of the building and show how we constrain the model by OCL-like rules to support a sound specification which can be matched against monitoring results afterwards. The paper presents our domain-specific language for modeling buildings and technical facilities that is implemented in a software-based tool used by domain experts and thus hopefully providing a suitable contribution to modeling the cyber-physical world.

## Categories and Subject Descriptors
H.1 [**Models and Principles**]: General; D.2 [**Software Engineering**]: Requirements/Specifications

## 1. INTRODUCTION
Currently buildings are equipped with a lot of technology that is rather not integrated and not very well connected to today's technological perspectives. In this paper we use an integrative approach from the cyber-physical systems initiative [4, 5, 11] to model buildings from the energy perspective. We introduce a domain-specific language [10] for modeling buildings and technical facilities focussing on improving their energy efficiency. Today's buildings and facilities are often equipped with sensors that produce a lot of operational data. On the basis of this sensor information a language is designed to model energetic properties of building as well as facility behavior. Additionally, the measured sensor data is used to analyze and to improve the energy efficiency.

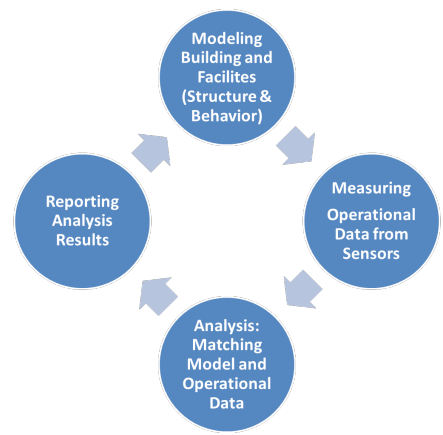

**Figure 1: Using models for a formalized planning and analysis process**

The language provides structural elements to model the real building context. Additionally behavioral elements are used to model the facilities' behavior and energetic properties. For modeling the behavior an adaption of the Object Constraint Language (OCL) [18] is used. In our scenario the OCL is not used in the context of UML modeling languages [19], like class diagrams or object diagrams, but in the context of buildings and their sensors. From this formalized notation, analysis algorithms are derived and executed dur-ing runtime. With this approach, we demonstrate that the use of software modeling languages can be adapted rather easily to other technical contexts by adapting the syntactical shape and the semantics with regards to the real world, but keep the internal modeling techniques as developed originally. With this approach we demonstrate that it is perfectly possible to use standard modeling techniques, e.g. from UML [19] or SysML [20] and adapt them as domain-specific languages for cyber-physical systems. The language was designed in cooperation with experts from the building and facility engineering domain.

## 2. THE ENERGY NAVIGATOR

The aforementioned features are all integrated in a software product called "Energie Navigator". The Energy Navigator[1] is a software-based tool to plan, measure and analyze the energy efficiency of buildings and technical facilities that are equipped with building management systems.

Today the common way to analyze buildings and facilities is part of the domain of monitoring which mainly involves collecting operational data from a building management system. After capturing the operational data of a building, experts are able to investigate the data using visualization tools. However this approach is not very effective, since the success depends on the expert's experiences. The know-how cannot be transferred to other buildings easily and since the analysis is done manually it cannot be repeated automatically. Especially after some optimization work on the buildings and their operation systems manual work has to be done once again. The monitoring approach does not include a closed loop, as shown in figure 1. But such a closed loop is a crucial aspect for a successful, sustainable and effective analysis process. The monitoring approach includes measuring, analysis and sometimes reporting functionalities but lacks methods of modeling buildings and feeding back gathered information into the specification.

To overcome this issue and close the loop the Energy Navigator provides the concept of a domain-specific language that can be used by domain experts to define set points. These desired operation values can then be matched with the operational data that is measured by the sensors. With the Energy Navigator experts can start modeling a building in the planning phase and the created model can then be used for implementing the building management system. Afterwards, the analysis is done automatically and the results are visualized in a comprehensive way. The expert knowledge can later be replicated for further buildings and the optimization loop is closed, since the effects of optimization tasks are detected easily.

The developed domain-specific language is based on the context of sensor information. This language is used by domain users, who are energy experts and facility managers, to model their real buildings. By this formalized description of the real world the buildings are instrumented to automatically analyze their energetic state.

The Energy Navigator has been developed in cooperation with domain experts from the domain of energy efficiency.

---
[1]http://synavision.de/demo/

During this cooperation the above described problems were identified and solution approaches integrated into the software. The software-tool is currently used by domain experts in several pilot projects applying our domain-specific language in real world projects that model energy efficient buildings.

The Energy Navigator is closing the optimization loop by introducing the specification as part of the design of an energy efficient building. The quality is improved by connecting measured sensor data with analyses algorithms and reporting functionality. The analyses results can be used for adapting the building operation parameters. Thus additional value is created.

## 3. DOMAIN MODEL

Our domain-specific language to model buildings and behavior of elements in buildings is based on the domain elements shown in figure 2. The root element represents a physical `Location`, like a building on company premises. A composite pattern is used for sublocations and subfacilities. `Facilities` are smaller subsystems inside a location, like a central heating unit. These are explained in more detail in 3.2.

To describe the functional behavior of the facilities constraint rules can be nested inside a facility or a location. `Rules` that describe the overall behavior of the location or model cause-effect correlations between multiple facilities are associated with a location, while rules that constrain the behavior of a single facility are nested inside a facility element. The rule elements use an embedded expression language to specify constraints over the other elements. These are explained in 3.3 in more detail. To get an idea of such a constraint rule, a simple room temperature constraint could be written as:

$$17 < Office.FirstFloor.Hall.roomTemperature \leq 25$$

with *roomTemperature* being the name of the temperature sensor installed in the hall of the first floor of the modeled office building.

Furthermore, additional location elements can be aggregated inside each other to be able to model hierarchical structures like departments, inside a single building. To model facilities more fine grained nesting of subfacilities inside facilities is also supported. Apart from those elements `Sensor`s with an associated `Unit` and associated `Values` can be used within the domain-specific language.

A sensor represents a physical sensor that is able to meter different kinds of values and is typically connected to a building automation system. The building automation system uses unique addresses for each sensor. This address is stored in the corresponding field of the sensor element. Since the available sensors are very heterogeneous they are able to meter all kinds of data having different units. Thus, each sensor has an associated unit. Storing the unit explicitly also enables automated unit conversion inside the business logic of the system.

The values of a sensor are stored as a tuple consisting of a timestamp and the actual value. The timestamp is given by the building management system and represents the time the building management system has triggered the sensor. Data values are usually triggered in fixed intervals, e.g. every 15 minutes. The sensors are always nested inside a location which serves as a container. Other elements like rules or facilities only use object references to the sensors stored in the location. Figure 2 represents a snippet of the overall model of the domain-specific language. Since we are dealing with a huge amount of values the Energy Navigator uses a cloud-based processing and high-performance database back-end.

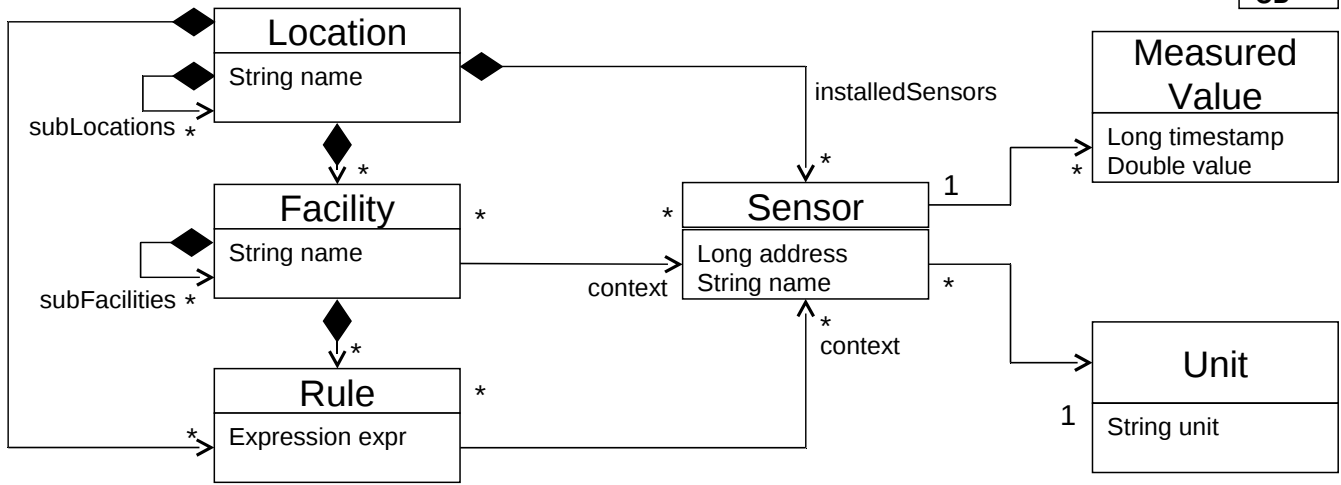

Figure 2: Domain model as class diagram

Other elements, presented in [9], are time routines and functions, being expressions like rules. In contrast to rules that evaluate to boolean values, functions evaluate to numerical values. Additionally constants, characteristics, and states [8] that model the different modes of a facility, can be used as elements of the domain-specific language. Elements can be referenced from other elements to enable reuse.

### 3.1 Sensors

Sensor elements represent the physical sensors that are installed inside and outside of a building. Each sensor has an address that maps to the physical address of the real sensor. In modern buildings there are often up to 3000 sensors installed, that are poorly documented. In the Energy Navigator each sensor element can be enriched with meta data. Besides a unique address, it is also possible to add a description, category and value type of a sensor. This information is used to handle values that are measured by this sensor in the building. Often the address of a sensor is a numeric value or a mixture of characters and numbers, e.g. the sensor for the room temperature in a specific office room of a building could have the address `EGS_GLT001a`. Without additional information interpreting the measured values would be impossible. However, with the description `Temperature room 001`, and the value type `double`, the context of the values is given more clearly. Additional attributes are available in the sensor. There are e.g. first value and last value as timestamp, that specify the timestamp of the first and last measured value of the sensor.

Each sensor belongs to one location, that is, the building where the real sensor is located at. In each building the address of a sensor is unique. Therefore the sensor can only be analyzed in the context of a location.

### 3.2 Facilities

In normal buildings sensors are used to measure and/or control the operation of facilities. Each facility consists of a name, description, a graphical representation and a context of all relevant sensors for this facility. Typical facilities of a building are hot water circuits or central air handling units. The diagram in figure 3 shows the functional design of a facility, with its pipes, sensors, and devices. These diagrams are created during the design of the building and can be reused in the Energy Navigator.

Each of theses facilities have already build in sensors or need additional sensors within a building in order to be functional. Because of the unique addresses within one building the sensor belongs to the location but can be associated to a facility. Additionally, several facilities can use one sensor for operation, e.g. a hot water circuit and a central air handling unit need the outdoor temperature sensor.

All sensors that are relevant for a facility are connected via the context of this facility. Only object references to the sensor are used. To assist the domain experts defining the relevant context the imported diagram of the facilities is used. In figure 3 a facility with its diagram and sensor context is shown. A domain expert can add concrete sensors from the building to the diagram and place them on top. The graphical representation of the facility consists of two parts. The first one is the circuit diagram that is imported, the second one is a layer for the concrete sensors from the building which are placed on top. Only relevant sensors for the facility need to be added to the context. Other sensors of the facility can be left out.

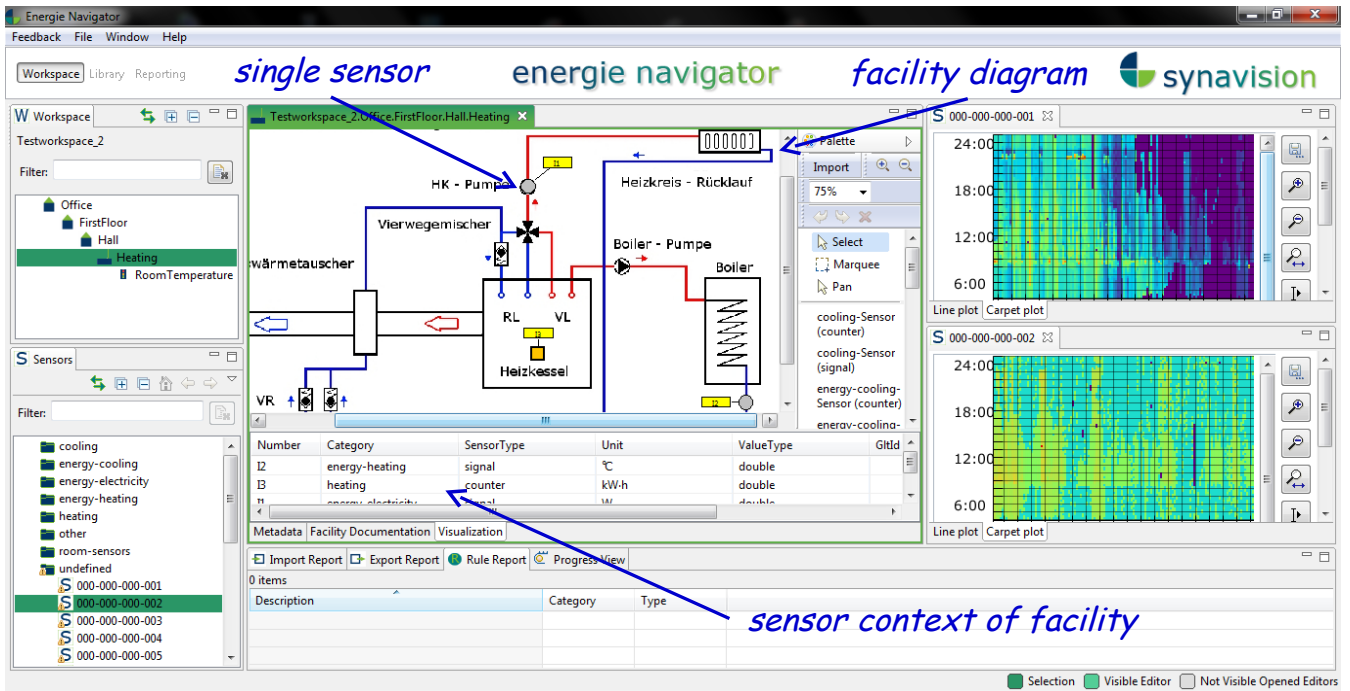

Figure 3: Facility with diagram and sensor context

## 3.3 Rules

Rules are elements of the domain-specific language of the Energy Navigator. A rule contains an expression written in a defined expression language. Its grammar has been defined using the MontiCore framework [16] that enables creating textual domain-specific languages by specifying abstract as well as concrete syntax in a single grammar file. Furthermore, it offers support for language embedding, inheritance and composition as well as context condition checking and editing support for concrete models.

The idea and concept behind the rules is based on the OCL. OCL conditions are mostly specified on the class diagram level and support expressing constraints over object graphs and objects. Analogously to OCL conditions rules are specified on the building specification level. This specification includes the elements of the domain-specific language and constrains their properties. As long as there are no values associated to a sensor the elements can't be evaluated and have no properties to constrain. So the rule is specified on the level of the elements and constrains the values of the sensors and the elements. Since the rule expression language is used by domain experts the idea of constraining instances, adapted from the OCL, is used but not all operators are supported. The expression language offers several operators like `+,-,*,/,>,≥,<,≤,→, ⇔, ∨,∧, if-then-else`. Rules and functions use the same expression language as its specification. Additional context conditions ensure that expressions in functions always evaluate to a numeric value while expressions in rules evaluate to boolean values.

Apart from the operators a reference concept can be used to address other elements. The elements and sensors are always located inside a context and thus can be addressed via their names and the path defined by the names of the context elements. To specify this we use full qualified names for the elements analogously to the full qualified names of types in Java. We can also address elements by using their simple names if they are in the same context, because the names are unique within their context. Since the data values are stored in discrete time slices we can compare the same slices for some elements of the domain-specific language at a given point in time. Depending, e.g. on the context of the rule from the example above, using the simple name would also suffice if the rule is specified inside the "hall" location. As shown in figure 4, this rule would evaluate the set of values associated with the sensor. For every value representing a single time slice a corresponding boolean value inside a new set of values would be created. The two constants do not have a connection to a specific point in time but are used to compare to every time slice. Thus, the information at which points in time the constraint was satisfied or not can be captured by the rule. The resulting values can then be visualized by using a carpet plot. A carpet plot displays a two-dimensional cartesian coordinate system with the hour of day shown on the ordinate and the day shown on the abscissa. Every time slice is then displayed as a colored point in the coordinate system. The color normally ranges from dark blue to dark red, based on the corresponding value. For boolean results the values are interpreted as green and red colors. Carpet plots are well suited for visualizing patterns or creating an overview of a large amount of data. More details on the carpet plot are provided in [9]. Furthermore there are several context conditions to be checked. Some context conditions are checked at design time of the rule, e.g. like checking whether the room temperature sensor has valid values. The specification of a rule using two different elements can also be checked at design time if the associated units match. Apart from that some context conditions can only be checked during monitoring where the sensors have actual data. Checking if the available time slices match, is an example for such a context condition, since the sensors can have different values with different timestamps associated. The domain expert is made aware of this by using warnings and errors in the front-end together with hints on what went wrong. The rule expression language also supports comparing past time slices by using an `@pre` like operator that can also be parameterized to jump back in time, e.g. to the previous time slice or the according time slice of the previous year.

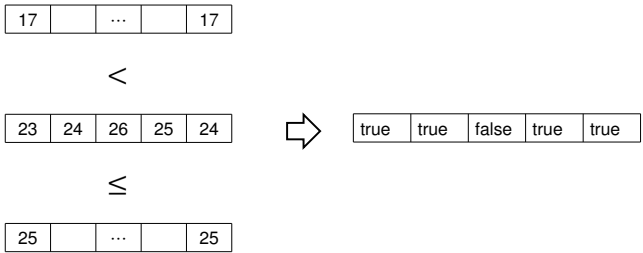

Figure 4: Evaluation of the example rule. Each time slice is evaluated and a new resulting time slice is created containing a boolean value.

## 4. RELATED WORK

There are several approaches to define a data model for buildings. One example is the Building Information Model (BIM) [6]. The BIM contains only static data describing the structure of a building. The data contained in the BIM is very fine granular and not focused on building optimization. It contains information about the different materials used to construct the building, information about the used doors, windows etc. and information about the last audit and renovation processes. It is designed to capture meta information of a building and not to model correlations and constraints inside a building. Thus, the information from the BIM could enrich our approach to model energy efficient buildings by taking additional meta information into account but is focused on another application domain. Furthermore, it is very prominent in rendering design models and virtual walkthroughs for planned buildings. ArchiCAD is a commercial software based on the BIM enabling rendering of virtual models [14]. In contrast to the Energy Navigator these approaches do not support automated analyses.

In [21] a transformation language is proposed that enables interoperability between different kinds of existing models and modeling languages. By creating a bridge between the STEP (Standard for the Exchange of Product model data) ISO standard [15] and the Eclipse Modeling Framework they are able to integrate these models into a common modeling language. Based on existing specifications this approach focuses on existing models and not on the specification of a domain-specific language for the the specification itself. This approach also just focuses on static data and not on dynamic analyses.

In [17] information systems for monitoring and managing energy efficiency of buildings are classified into four groups. They are divided into Energy Information Systems (EIS), Demand Response Systems (DRS), Enterprise Energy Management (EEM) and Web-Based Energy Management and Control Systems Classification (Web-EMCS). All of them are able to gather, aggregate and display data. DRS focus on the communication between energy providers and customers, EEM focuses on enabling benchmarking and optimizing complete business enterprises with different sites and creating management reports for financial analysis. Apart from that Web-EMCS focus on having a single application server and a database server that different user groups can access. The application server is able to connect to several buildings and control the buildings' behavior and is also able to aggregate and visualize metered data by querying the database server. [17] also provides a list of tools which are categorized according to these categories. It shows that most tools can be categorized into either EIS or Web-EMCS and offer different possibilities for collecting data in different intervals or visualizing it in different ways. The comparison between actual and desired behavior as supported by the Energy Navigator is missing.

In [12] a study of different commercial products has been conducted. The study shows that most commercial products focus on the data itself. The tools are able to collect it from different sources and are also able to display it via different front-ends tailored to a specific user group.

In [13] best practices, common measurements and advices on which measurement to apply on which facility are presented. Since the output of a EIS is used as a discussion basis common conversion factors that should be used within the monitoring process are introduced. These conversion factors represent formulas that e.g. convert the gas flow into the energy demand. These conversions that should be applied according to the best practices can directly be modeled with our concept of rules and functions. Best practices for the resolution of the data values are also given in [13].

Following [13] the energy manager should have intense knowledge about a lot of different conversion methods and formulas to get anything useful from the measured data. In addition to this the DIN EN ISO 16484 [7] also suggests a description of the functional behavior of a system in a state-oriented way. Within the Energy Navigator this can be modeled by our concept of states .

Another research area is the simulation of the energy performance of buildings. During the planning or optimization process different operation parameters can be varied to simulate effects. Tools like EnergyPlus[2], Ecotect[1] and eQuest[3] can be used to support the planning process. Compared to the Energy Navigator a simulation does not offer a concrete specification of the desired building behavior that can be used for validation during operation.

To sum it up, there are a lot of existing commercial monitoring tools. Most of these tools only focus on monitoring and thus on presenting the metered data in an aggregated visual way. They usually work in a data point oriented way that offers possibilities to filter some data points, add some meta data and visualize multiple data points in one diagram.

It is not possible to model the buildings according to their actual layout and add some meaning to the data points. It is also not possible to constrain the specified models as it is with our domain specific language for modeling buildings and facilities.

## 5. CONCLUSION

We have presented our approach to model cyber-physical systems, namely buildings and facilities, from an energetic point of view. We have presented our domain-specific language for modeling physical buildings and facilities. We presented sensors, locations, facilities and rules as chosen elements of the language. While sensors are associated with time discrete values, facilities and locations create a context for rules constraining the specification of the building. We have also presented the OCL-based idea of the expression language used within the rules. The specification of a building acts as the underlying model while the rule itself constrains instances of the specification. We have shown that our approach abstracts from a data point oriented view. It is not closed in to a fixed interval because it can be configured by the user and it supports conversion of different data like proposed in the best practices in [13] with the abstract modeling concept of the expression language used in rules or functions. With this approach we adapted widely used modeling languages to the domain of energy efficient buildings and used it to model cyber-physical systems. At the moment, the Energy Navigator is a tool that closes the loop between the specification of buildings and facilities, enables reporting, like financial reports or maintenance reports, as well as verification and visualization. Verification can be achieved through metered data but a control channel directly feeding the aggregated information back to the building is currently not a part of the Energy Navigator. By implementing such a control channel automated counter actions for saving energy are possible to be applied to the building, like e.g. shutting the blinder instead of increasing the air conditioning. In addition, the visualization of the data helps to find patterns in the data and to find correlations, e.g. by using the carpet plot. So far, the Energy Navigator has been used by domain experts in several pilot projects. First experiences show that the Energy Navigator can help to fill the gap between planning and operation of buildings and to improve the energy efficiency.